# Magnetovolume effect in $SrRu_{1-x}Co_xO_3$ ($x$ = 0.0, 0.05)

**M. Manikandan and R. Mahendiran***

*Department of Physics, National University of Singapore, 2 Science Drive 3,*
*Singapore 117551, Republic of Singapore*

## Abstract

Polycrystalline $SrRu_{1-x}Co_xO_3$ ($x$ = 0.0, 0.05) and $CaRuO_3$ were rapidly synthesized (< 1 hour) by microwave irradiation of oxide powders, and their magnetic, magnetoresistance, thermal expansion, and magnetostriction properties were investigated. The microwave-synthesised $SrRuO_3$ exhibits a ferromagnetic transition at $T_C$ = 160 K, metallic-type resistivity, and negative magnetoresistance, with magnitudes comparable to those of a sample synthesized by conventional heating over 24 hours. Upon lowering the temperature from 300 K, the linear thermal expansion shows a transition from the usual contraction in the paramagnetic state to spontaneous expansion in the ferromagnetic state (invar-like effect). The application of an external magnetic field at a fixed temperature results in isotropic expansion of the length, implying a positive magnetovolume effect. The volume magnetostriction is $\omega$ = 40 ppm at 10 K in a magnetic field of 50 kOe, and it reaches a maximum value ($\omega$ = 60 ppm) close to $T_C$. The spontaneous thermal expansion is diminished in $SrRu_{0.95}Co_{0.05}O_3$. While the magnetostriction is anisotropic at 10 K, the isotropic behaviour is recovered above 80 K, and the maximum value of the positive magnetovolume is comparable to that of the parent compound. Our results suggest that the magnetovolume effect in $SrRu_{1-x}Co_xO_3$ is related to competition between robust tilting/rotation of $RuO_6$ octahedra and spin-orbit interaction of the doped $Co^{2+}$ ions.

**Corresponding author E-mail: phyrm@nus.edu.sg*

## 1. Introduction

$SrRuO_3$ is the only itinerant electron ferromagnetic metal among 4*d*- perovskite oxides. It has Curie temperature of $T_C$ = 162 ± 3 K and crystallizes in an orthorhombic phase at room temperature.[1,2] Due to its high electrical conductivity, excellent lattice matching with many other perovskite oxides and good chemical stability, it is widely used as a spacer layer in oxide-based magnetic tunnel junctions and as a conducting electrode in resistive/ferroelectric devices.[3,4,5,6,7] While magnetic and galvanomagnetic properties have been extensively explored in single crystals,[8,9] thin films and polycrystalline samples,[10,11,12,13,14,15] less is known about their magnetoelastic properties. Analysis of temperature dependent powder X-ray[16] and neutron diffraction data[17,18,19] for powdered $SrRuO_3$ reveals that the crystal structure (orthorhombic, *Pnma* space group) remains unchanged as the temperature decreases from 300 K to 2 K but the unit cell volume transitions from normal contraction in the paramagnetic state to nearly temperature independent behavior in the ferromagnetic state, reminiscent of the Invar effect found in the $Fe_{0.65}Ni_{0.35}$ metallic alloy.[20,16] However, the isostructural $CaRuO_3$ remains paramagnetic down to 4.2 K and exhibits a normal volume contraction (positive thermal expansion) with decreasing temperature. Interestingly, the invar effect is induced by Fe doping in $CaRu_{0.85}Fe_{0.15}O_3$[21] but weakened in Cr-doped $SrRu_{0.87}Cr_{0.13}O_3$.[22] Therefore, it is of fundamental interest to investigate the stability of the invar effect in these compounds under an external magnetic field. There is only one report available on the influence of external fields on the physical dimensions of $SrRuO_3$. Kukenmöller *et al.*[23] reported the magnetic field dependence of fractional length change (magnetostriction, $\lambda$) in $SrRuO_3$ single crystal. The magnetostriction measured at 5 K along [101] axis showed a sudden jump ($\lambda_{101}$= 6 x $10^{-3}$) around $H$ = 15 kOe in the first field sweep but a much smaller change and a non-abrupt behavior while reversing the field to zero. On the other hand, magnetostriction along the [010] axis was negligible. Furthermore, linear thermal expansion measured in the zero field during warming depends on the sample's previous magnetic history (magnetic memory effect). Magnetic field-induced twin boundary reorientation was suggested as a possible origin for this memory effect.

In this context, magnetostriction in polycrystalline $SrRuO_3$ is of interest, but no reports are available to date. The two main aims of this paper are: 1. Rapid synthesis of $SrRuO_3$ by microwave (MW) irradiation, which has not been reported so far 2. To investigate magnetism, electron transport, magnetoresistance, and magnetostriction in the MW-irradiated sample and compare them with those of a sample synthesised by conventional heating (CH) in an electrical furnace, following the standard solid-state reaction route. The conventional synthesis of bulk $SrRuO_3$ involves mixing $SrCO_3$ and $RuO_2$ powders in a stoichiometric ratio, grinding, and calcining at high temperatures (900 to 1300 °C) multiple times, then pressing into pellets followed by sintering the compressed pellets in an electrical furnace. Calcination and sintering processes take 48 to 72 hours in the CH method. In this work, both synthesis and sintering were carried out in a fraction of time (< 1 hour) using microwave (MW) irradiation. The chemical reaction in mixed powders of $SrCO_3$ and $RuO_2$ is initiated by absorption of microwave radiation primarily by $RuO_2$, since $SrCO_3$ is a poor microwave absorber. When electrical dipoles and/or conduction electrons of $RuO_2$ are unable to oscillate in-phase with the electric field of the impinging microwaves, the absorbed MW energy is dissipated as heat internally in $RuO_2$ which accelerates the rate of diffusion of ions between $RuO_2$ and $SrCO_3$ leading to formation of the desired compound in a shorter time compared to the CH method. Some excellent reviews highlight the advantages of microwave synthesis or heating (MWH) for different materials.[24,25,26] MWH has been used to synthesize pure and divalent cation doped $LaMO_3$ (M= Mn, Co, Cr) using metal hydrates or chlorides as starting reactants because they are good microwave absorbers[27,28,29,30,31], but MW synthesis of $SrRuO_3$ directly from an oxide precursor has not been reported so far. Additionally, less is known about the impact of MWH on the electrical and magnetic properties compared to those obtained via the CH method.

High-purity $SrCO_3$, $RuO_2$ and $Co_3O_4$ powders were used as precursors. They were taken in a stoichiometric ratio and mixed uniformly using an agate mortar and pestle. An alumina crucible containing 5 grams of mixture was placed at the centre of a multi-mode microwave furnace (Milestone PYRO) and irradiated with microwaves (MW) at 1000 W and 2.45 GHz for 10 min to reach 800 °C. The powder was then soaked at 800 ºC for 20 min before switching off the MW power which allowed the powder to be air quenched from 800 ºC to room temperature in 100 min. The MW irradiated powder was reground homogeneously, pressed into a 12 mm diameter pellet, and irradiated again to reach and react at 900° C for 40 minutes, and then air quenched. Another five grams of powder was placed in a conventional electrical furnace and heated to 900 °C at the rate of 5 °C/min, soaked at 900 °C for 12 hours and then cooled to room temperature at 5 °C/min. The powder was re-ground homogeneously, pelletized, and sintered at 900 °C for another 12 hours with the same heating and cooling rates. A portion of the pellet was cut and crushed into powder for X-ray diffraction with Cu-$K\alpha_1$ (1.5406 Å) radiation. The magnetization, four-probe electrical resistivity, and magnetoresistance were measured using a Physical Property Measurements System (PPMS) with appropriate probes. Magnetoresistance (*MR*) is defined as $MR = [(\rho(H) - \rho(H = 0)]/\rho(H = 0)$ where $H$ is the magnetic field and $\rho$ is the resistivity. Thermal expansion and magnetostriction of a cube-shaped (2 mm x 2mm x 2mm) sample cut from a larger pellet were measured using a capacitance dilatometer probe attached to the PPMS. The capacitance change of the dilatometer was measured using a high-resolution capacitance bridge (Andeen-Hagerling, model AH2500A). Magnetostriction is defined as $\lambda = (L(H)-L(0))/L(0)$ where $L$(H) and $L(0)$ are the lengths of the sample in a magnetic field $H$ and at $H = 0$ kOe.

## 2. Results and Discussion

Fig. 1 shows the powder X-ray diffraction pattern (black circles) of MWH-$SrRuO_3$ with the Rietveld fit (red line) and the difference between the fit and the original data (blue line). The diffraction peaks are consistent with an orthorhombic structure (space group *Pnma*). The refined lattice parameters are: $a = 5.5379(2)$ Å, $b = 7.8461(3)$ Å and $c = 5.5631(2)$.

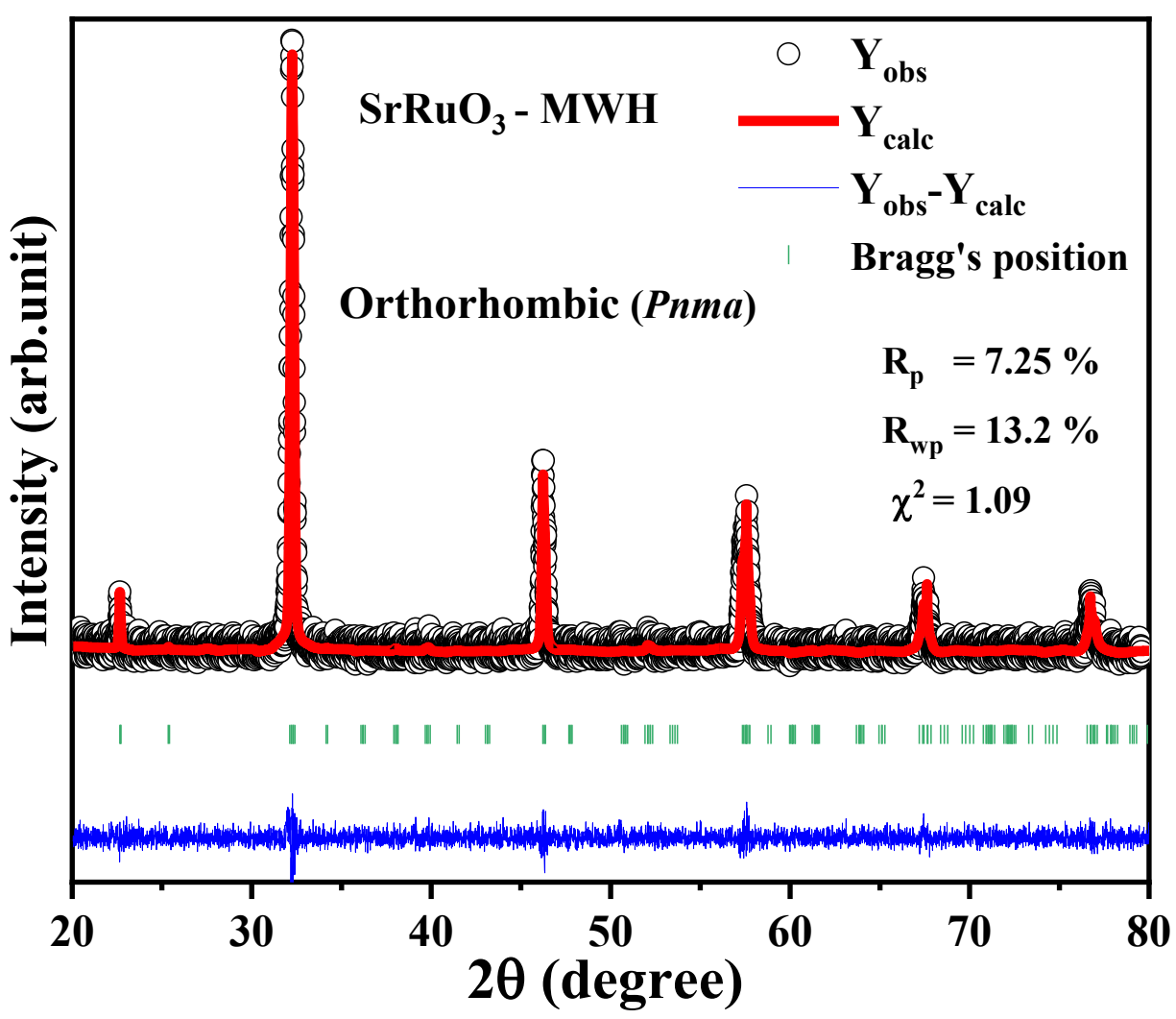

**Fig. 1**. Rietveld refined X-ray diffraction pattern of the microwave-synthesized $SrRuO_3$.

Fig. 2 illustrates the temperature dependences of (a) magnetization $M(T)$ in a field of $H = 1$ kOe, (b) *dc* resistivity $\rho(T)$, and (c) linear thermal expansion ($\Delta l/l_0$) measured in zero external magnetic field, where $l_0$ is the length of the sample at 300 K. The rapid increase of $M(T)$ around 160 K signals the onset of ferromagnetic ordering. The ferromagnetic Curie temperature identified from the inflection point of $dM/dT$ is $T_C = 160$ K which is consistent with the available data on polycrystalline $SrRuO_3$ synthesized by the CH method.[10,14,16,18] $\rho(T)$ decreases with decreasing temperature (metallic behavior) in both the paramagnetic and the ferromagnetic states, but with a noticeable change in slope around $T_C$. The paramagnetic to ferromagnetic transition is accompanied by a transition from usual thermal contraction of length in the paramagnetic state to anomalous expansion (increase in length) below $T_C$ (see Fig. 2(c)). The volume contraction due to anharmonic phonon contributions is overcompensated by an anomalous magnetic contribution as the temperature decreases below $T_C$. According to earlier X-ray diffraction studies,[16,18] the *a*-lattice parameter in the orthorhombic–symmetry shows normal contraction and the *b*- parameter increases slightly, whereas the *c*- parameter is temperature independent below $T_C$. The temperature independence of the *c*-parameter dominates the temperature variation of the unit cell volume. The inset in (c) shows that the negative thermal expansion behavior seen in the ferromagnetic $SrRuO_3$ is absent in the isostructural $CaRuO_3$ which remains paramagnetic down to 4.2 K in agreement with X-ray diffraction studies.[16,18]

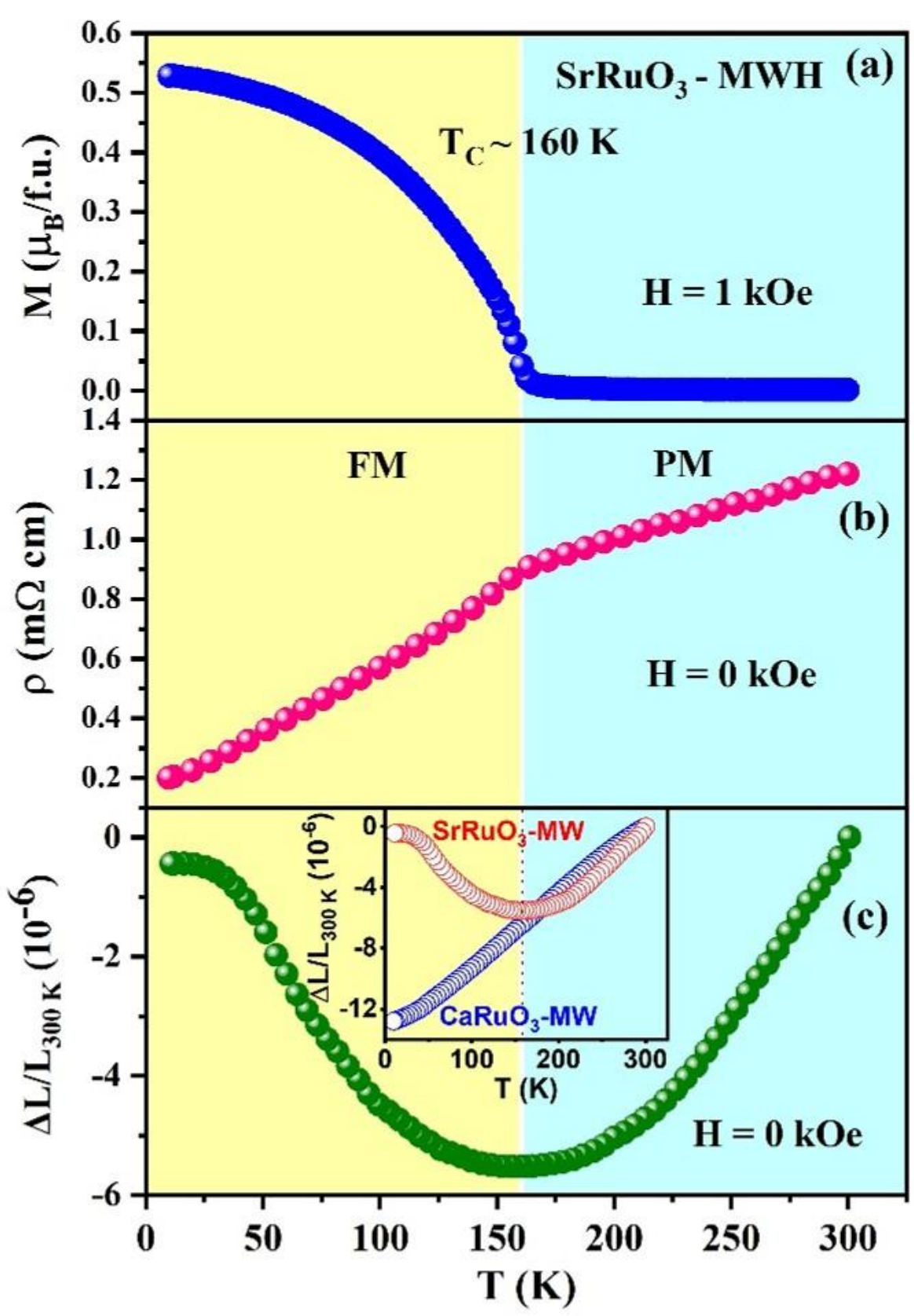

**Fig. 2.** Temperature dependence of **(a)** magnetization (*M*) under a magnetic field of $H = 1$ kOe **(b)** *dc* resistivity and **(c)** linear thermal expansion in zero external magnetic field in $SrRuO_3$ synthesized by microwave irradiation. The length of the sample at room temperature ($L_{300K}$) is taken as the reference. The ferromagnetic and paramagnetic regions are highlighted in different colours. The inset in **(c)** compares the linear thermal expansion in paramagnetic $CaRuO_3$ and ferromagnetic $SrRuO_3$.

The influence of an external magnetic field on the *dc* resistivity was studied during a temperature sweep at $H = 0$ and 50 kOe as well as by scanning the magnetic field at several fixed temperatures. The top-left inset of Fig. 3 illustrates the field dependence of magnetoresistance (*MR*) within the ferromagnetic state, and the bottom-right inset shows *MR* vs H for $T \geq T_C$. At 10 K, *MR* is positive at low fields, reaches a maximum around $H = \pm H_C$, becomes negative at $H > H_C$, and exhibits hysteresis extending up to 40 kOe. The hysteresis narrows with increasing temperature. In the paramagnetic state, *MR* is proportional to $-H^2$ over a wide field range, whereas in the ferromagnetic state, it increases linearly with the magnetic field well above the coercive field ($H_C$). The main panel shows *MR* for $H = 50$ kOe obtained from $\rho(T)$ measured under $H = 0$ and 50 kOe. The maximum value of *MR* is -2.25 % at $T_C$, which falls within the range of 2 to 6 % observed in single-crystalline and polycrystalline $SrRuO_3$.[11,15]

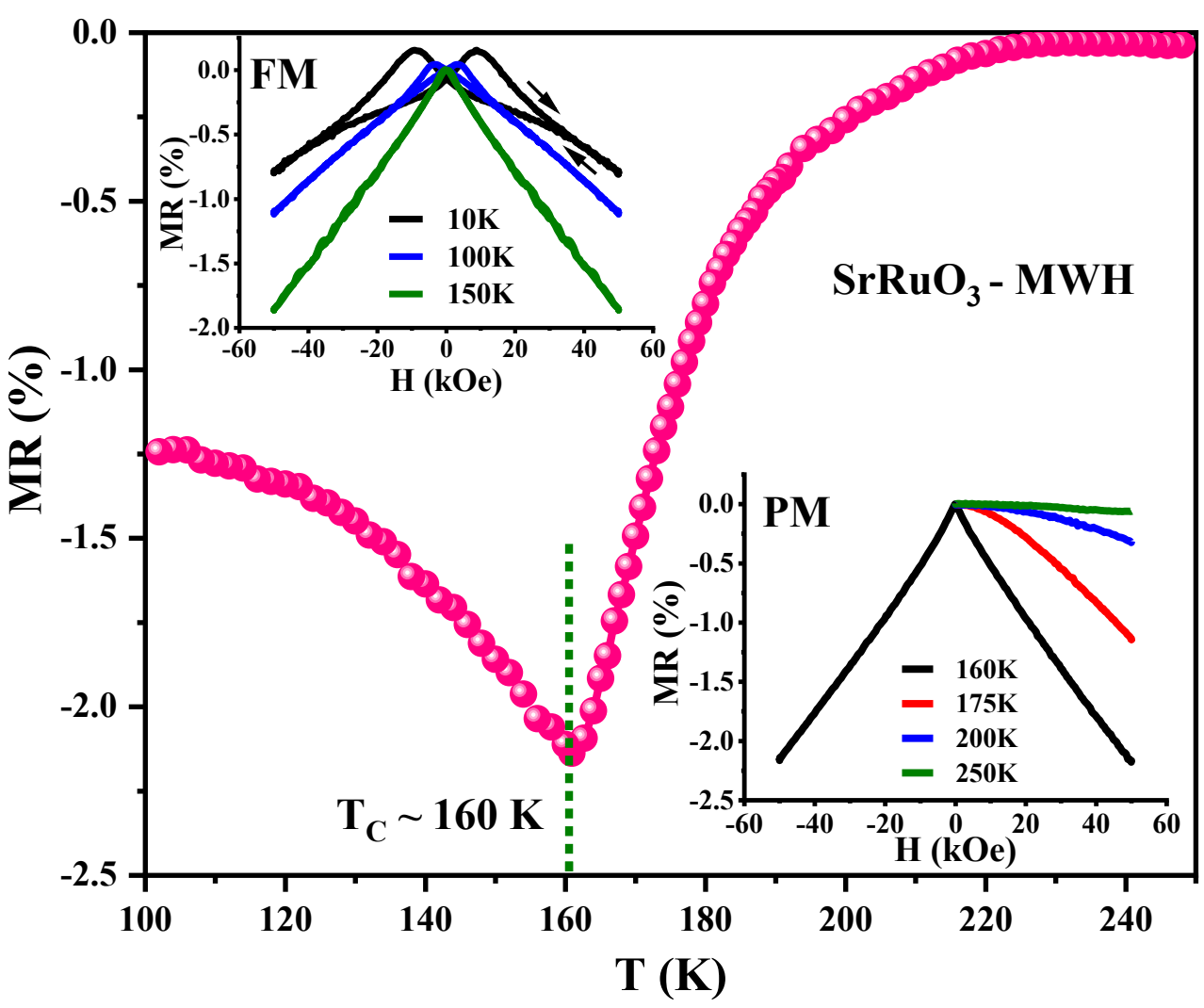

**Fig. 3.** Temperature dependence of magnetoresistance of MW-$SrRuO_3$ for $H = 50$ kOe obtained from the temperature sweep under 0 and 50 kOe. The insets show the field dependence of magnetoresistance in ferromagnetic **(top-left)** and paramagnetic regions **(bottom-right)**.

The field dependence of magnetization $M(H)$ at 10 K (Fig. 4(a)) shows a large coercive field ($H_C = \pm$ 10 kOe) with hysteresis extending up to 40 kOe. Magnetic saturation is not reached even at 50 kOe. Longo *et al.*[2] reported non-saturation of magnetization even at $H =$ 125 kOe in single-crystalline $SrRuO_3$. The lack of saturation suggests a large magnetocrystalline anisotropy. The observed magnetization of $0.91\mu_B$/f.u. at 50 kOe is lower than 2 $\mu_B$/f.u. expected for spin only value of $Ru^{4+}$: $t_{2g}^4$ ($S = 1$) in ionic model. A range of magnetic moments (~ 0.8-1.6 $\mu_B$/Ru) has been reported in the literature for $SrRuO_3$ depending on the synthesis conditions.[1-15]. The much reduced magnetic moment relative to the spin only value is attributed to strong hybridization of O-2*p* and Ru-4*d* orbitals and the itinerant character of the electrons.[32]

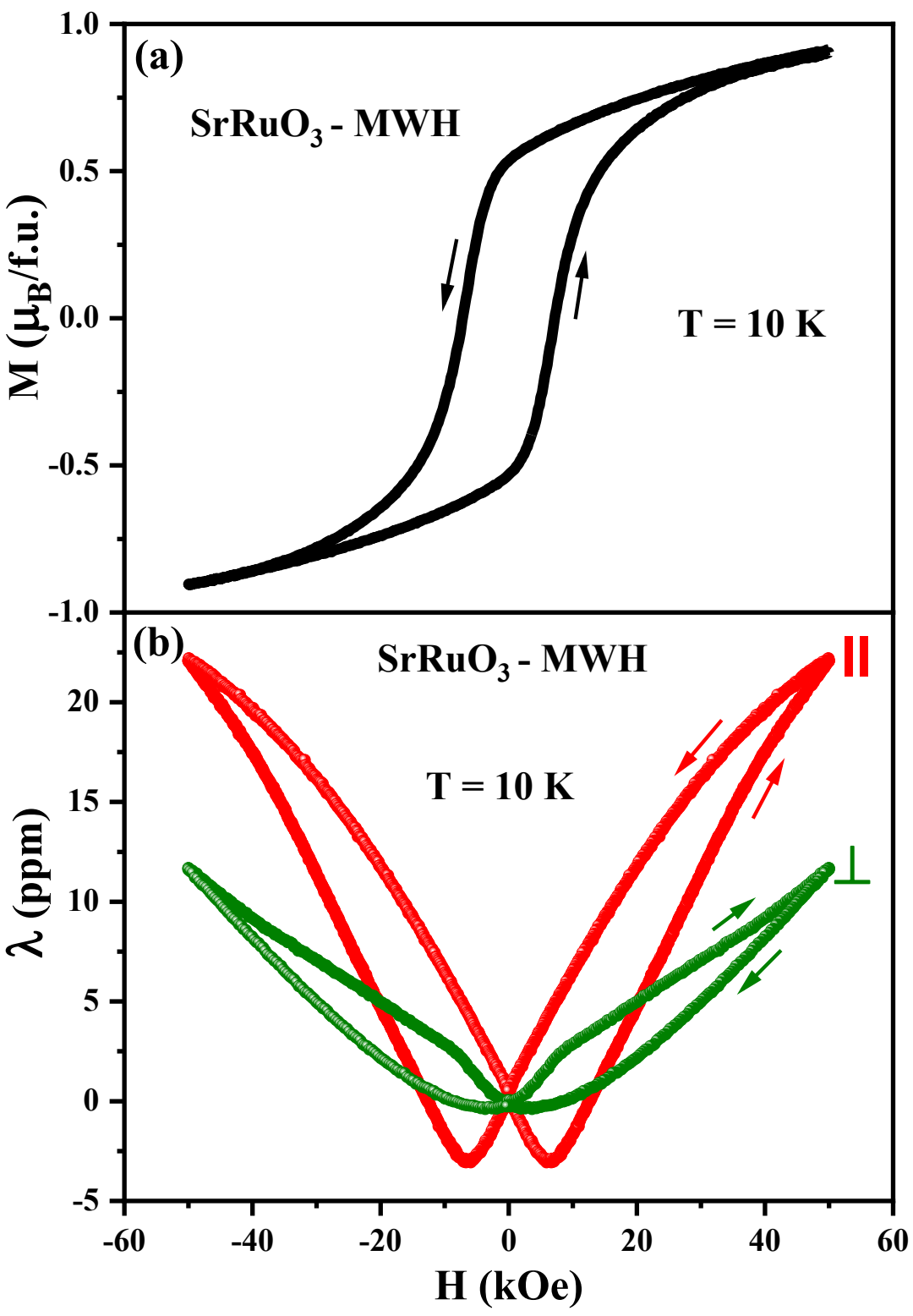

**Fig. 4.** Field dependence of **(a)** magnetization and **(b)** parallel ($\lambda_{par}$) and perpendicular ($\lambda_{per}$) magnetostriction of microwave synthesized $SrRuO_3$.

Fig. 4(b) illustrates the field dependence of parallel ($\lambda_{par}$) and perpendicular ($\lambda_{per}$) magnetostrictions of MWH-$SrRuO_3$ at 10 K. As the field is swept from +50 kOe to -50 kOe, $\lambda_{par}$ decreases and goes through a minimum around $H$ = -10 kOe, i.e., around the coercive field in the reverse field direction, and then increases without saturation up to -50 kOe. A similar trend appears when the field is scanned from -50 kOe to +50 kOe and it results in a hysteresis. The hysteresis in magnetostriction extends up to the maximum field, and it is much wider in $\lambda_{per}$ compared to $\lambda_{par}$. Interestingly, the magnetostriction is isotropic, i.e., the sign is positive for both $\lambda_{par}$ and $\lambda_{per}$, unlike a positive $\lambda_{par}$ and negative $\lambda_{per}$ observed in 3*d* ferromagnets: 30% Sr-doped $LaCoO_3$ and $LaMnO_3$.[33] The maximum value of $\lambda_{par}$ = 22 ppm is roughly twice that of $\lambda_{per}$ = 11 ppm at 10 K.

Figure 5(a) and (b) show the field dependence of $\lambda_{par}$ and $\lambda_{per}$ at $T$ = 20, 100, 140, and 180 K, respectively. The hysteresis in $\lambda_{par}$ decreases with increasing temperature, and it is absent at 180 K, where the sample is in the paramagnetic state. The hysteresis in $\lambda_{per}$ is abnormally high at 100 K compared with 20 K. The inset of Fig. 5(a) illustrates the temperature dependence of $\lambda_{par}$ and $\lambda_{per}$ extracted from the field sweep data at multiple temperatures. As the temperature rises from 10 K, $\lambda_{par}$ initially increases, passes through a broad maximum around 100 K and then decreases rapidly near $T_C$. By contrast, $\lambda_{per}$ exhibits a broad minimum and then peaks sharply just below $T_C$, before becoming negligibly small in the paramagnetic state. The volume magnetostriction ($\omega = \lambda_{par} + 2\lambda_{per}$) and the anisotropy magnetostriction ($\lambda_t = \lambda_{par} - \lambda_{per}$) were calculated and plotted in the inset of Fig. 5(b). At 10 K, $\omega$ = 40 ppm, and it gradually decreases until 150 K. Above 150 K, it increases rapidly and reaches peak of 60 ppm around 160 K before falling to negligibly small values in the paramagnetic state. However, $\lambda_t$ increases with temperature, reaches a maximum around 100 K and then decreases rapidly as $T_C$ is approached.

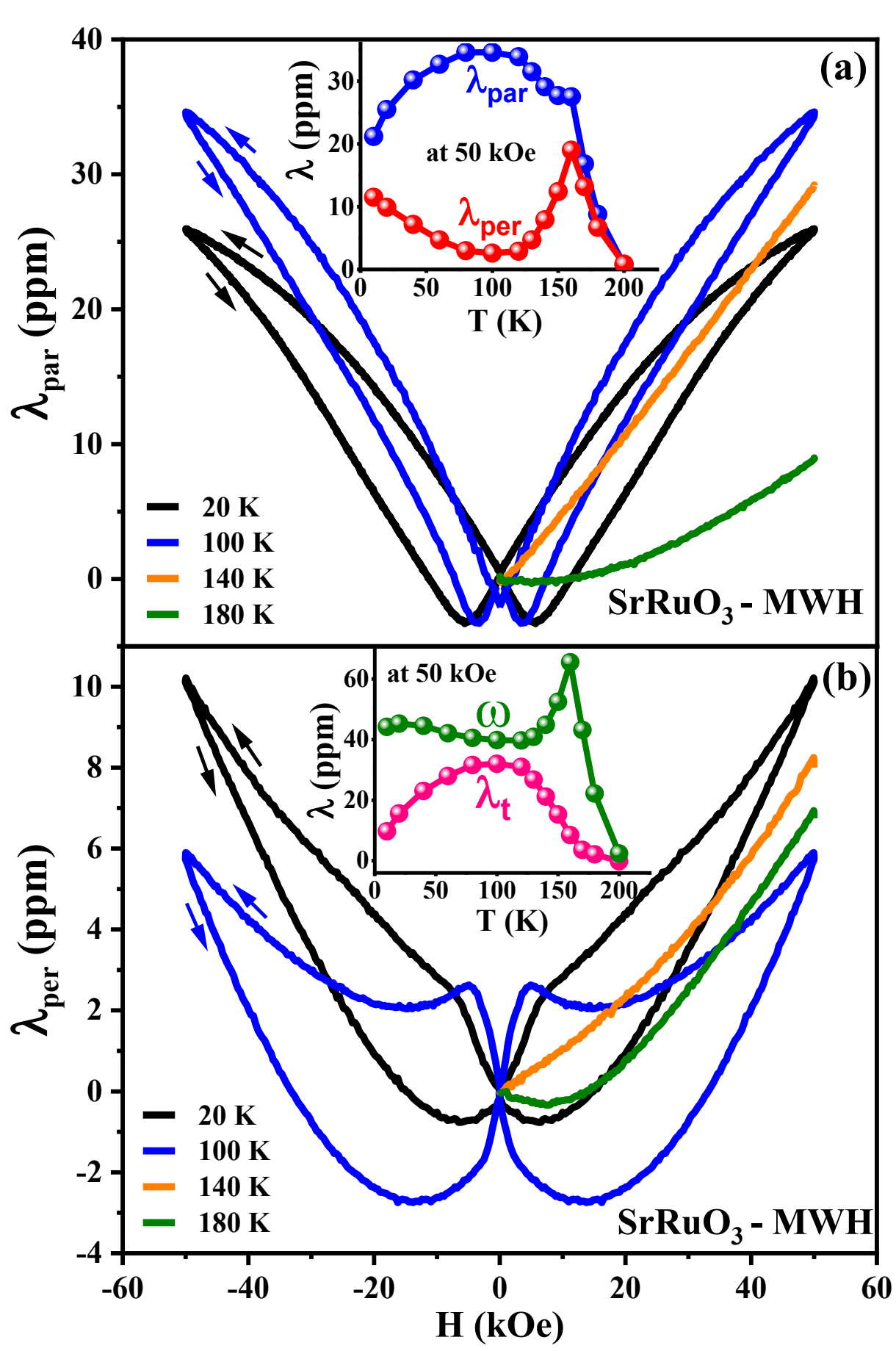

**Fig. 5.** Field dependence of **(a)** parallel ($\lambda_{par}$) and **(b)** perpendicular ($\lambda_{per}$) magnetostrictions in the microwave-synthesized $SrRuO_3$. Insets in **(a)** show the temperature dependence of $\lambda_{par}$

and $\lambda_{per}$ collected at $H = 50$ kOe from respective $\lambda(H)$ isotherms and in **(b)** show the calculated volume ($\omega$) and anisotropy ($\lambda_t$) magnetostrictions.

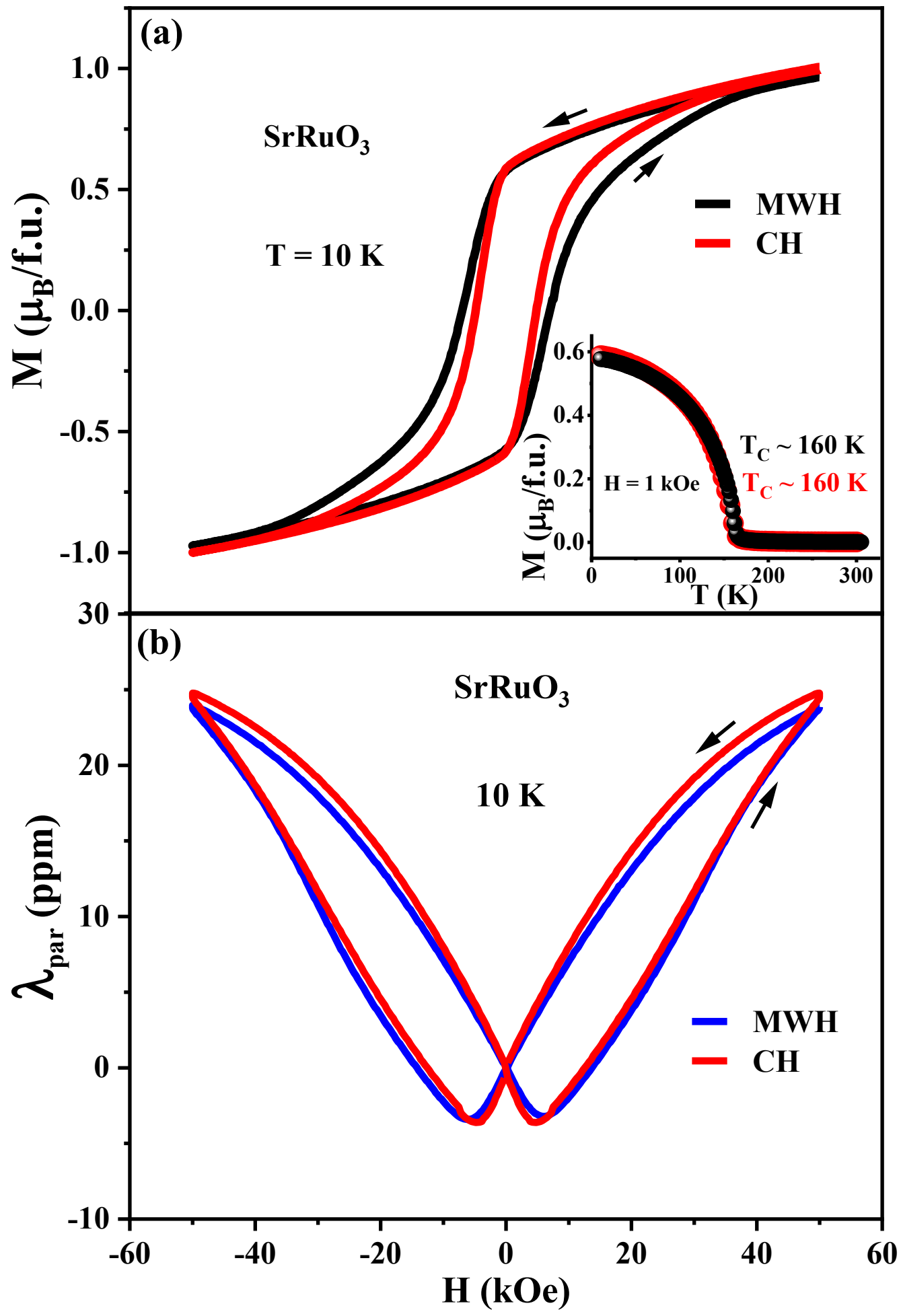

**Fig. 6**. Comparison of field dependences of **(a)** magnetization $M(H)$ **(b)** longitudinal magnetostriction ($\lambda_{par}$) at 10 K for the samples synthesized by microwave heating (MWH) in a microwave furnace and conventional heating (CH) in an electrical furnace. Inset in **(a)** compares the temperature dependence of magnetization under $H = 1$ kOe in both samples.

In order to verify the results obtained in the MWH sample, we prepared $SrRuO_3$ by conventional heating in an electrical furnace and compared its magnetization and magnetostriction results with those of MWH-$SrRuO_3$. Fig. 6(a) & 6(b) compare the field dependence of magnetization and longitudinal magnetostriction ($\lambda_{par}$) at 10 K for $SrRuO_3$ synthesized by MWH and CH methods, respectively. The inset in Fig. 6(a) shows that ferromagnetism sets in at approximately the same temperature ($T_C$ = 160 K) in both samples. The MWH-sample shows a higher coercive field ($H_C$ = 7.26 kOe) than the CH sample ($H_C$ = 4.624 kOe). The values of magnetization and magnetostriction at 50 kOe are comparable in both samples. This indicates that MW synthesis did not degrade magnetization and magnetostriction, despite a twenty-fold reduction in synthesis time.

We did not observe magnetostriction of the order of $10^{-3}$ reported by Kukenmöller *et al* in $SRruO_3$ single crystal during the initial field sweep, possibly because of the polycrystalline nature of our sample. The maximum value of volume magnetostriction (ω) in MWH-$SrRuO_3$ is only 60 x$10^{-6}$ near $T_C$, which is smaller than in ferromagnetic manganites, for-which $\omega$ is of the order of $10^{-3}$ around $T_C$.[31] The sign of $\omega$ is negative in doped manganites, unlike in $SrRuO_3$, indicating different mechanisms of the magnetovolume effect between these ferromagnets. In manganites that undergo a paramagnetic insulator-ferromagnetic metal transition, charge carriers are self-trapped as polarons (or magnetic polarons) by strong electron-phonon interaction in the paramagnetic state at zero magnetic field and the paramagnetic-ferromagnetic transition is accompanied by a spontaneous volume contraction in zero field. As the applied magnetic field increases the magnetization in the paramagnetic state, magnetic polarons expand in size and coalesce. This leads to delocalization of charge carriers, which in turn suppresses the excess volume (negative magnetovolume effect) around $T_C$ and in the paramagnetic state not far from $T_c$. However, electrons are itinerant in $SrRuO_3$ above and below $T_C$. It was suggested that the invar effect in $SrRuO_3$ is coupled with the ferromagnetic ordering of Ru moments via freezing of $RuO_6$ octahedra tilt about [100] axis.[18] Also, the in-plane <Ru-O> bond length shows an increase below $T_C$.[19] Our results indicate that the tilting/rotation of $RuO_6$ octahedra and <Ru-O> bond length possibly increase as domain magnetization rotates towards the field direction. In other words, the invar behavior is robust under magnetic fields at least up to 50 kOe.

R. Rajan *et al*.[34] reported an increase in $T_C$ to 190 K and a weakening of the invar effect in Cr doped $SrRu_{0.87}Cr_{0.13}O_3$. On the other hand, $T_C$ was found to decrease in Co-substituted $SrRu_{1-x}Co_xO_3$ ($x < 0.13$), however, the unit cell volume as a function of temperature was not studied.[35] The presence of $Co^{3+/}Co^{2+}$ and $Ru^{4+}/Ru^{5+}$ was detected by X-ray absorption spectroscopy. We have also synthesized $SrRu_{0.95}Co_{0.05}CoO_3$ by a microwave irradiation method and investigated magnetostriction in this compound. The main panel of Fig. 7 shows $M(T)$ under $H$ = 1 kOe on the left-y axis and zero field linear thermal expansion on the right y-axis. The $T_C$ of the sample is 150 K, which is 10 K lower than that of undoped $SrRuO_3$. The $M(H)$ graph shown in the inset indicates an enormous increase of $M$ at the maximum field, from 0.91 $\mu_B$/f.u. in the undoped $SrRuO_3$ to 1.5 $\mu_B$/f.u. in the Co-doped $SrRuO_3$. Interestingly, the thermal expansion anomaly seen in $SrRuO_3$ is much weakened in the ferromagnetic state of the Co- doped $SrRuO_3$.

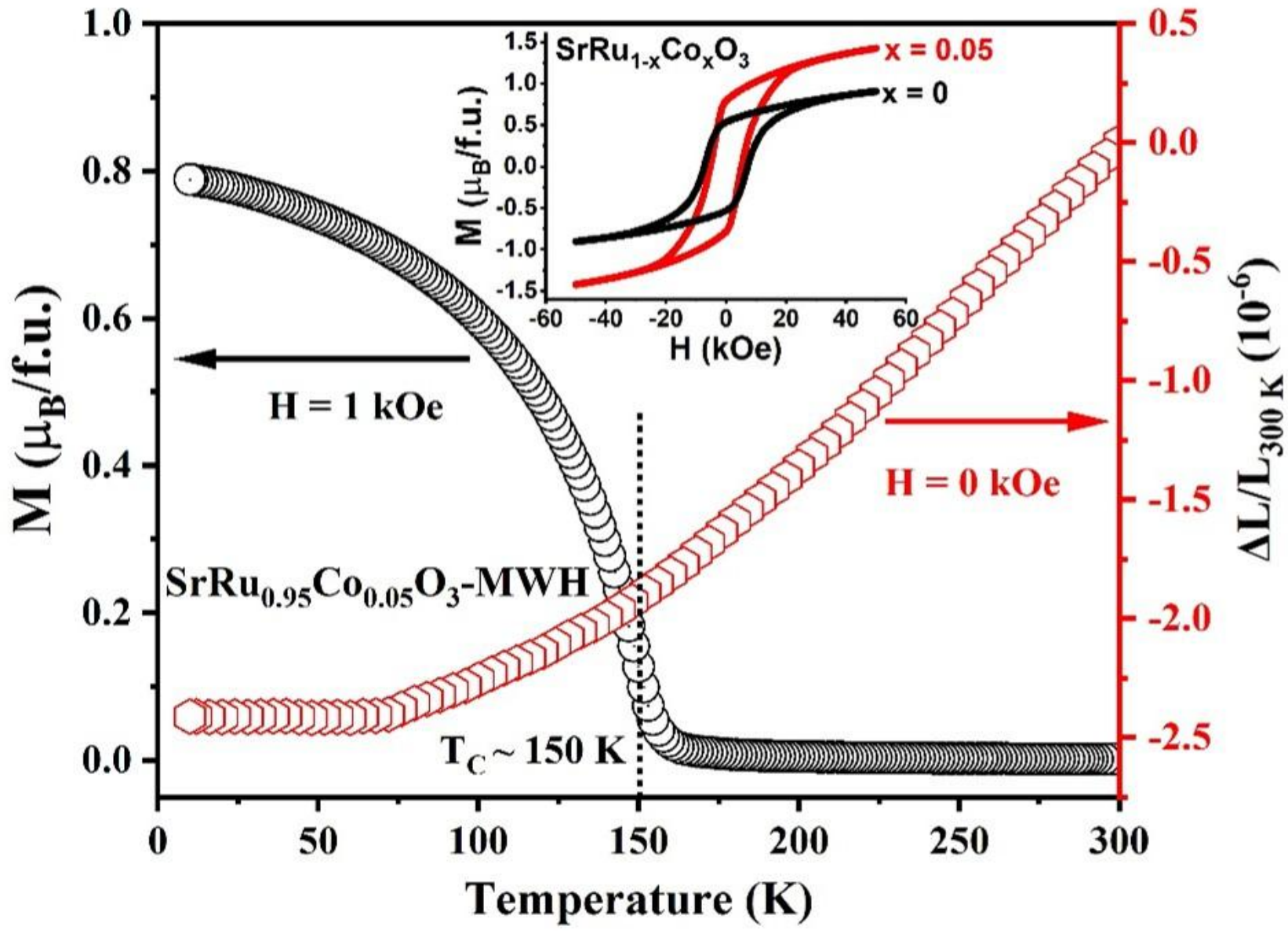

**Fig. 7.** Temperature dependence of magnetization $M(T)$ (left y-axis) and linear thermal expansion (right-y axis) of microwave-synthesized $SrRu_{0.95}Co_{0.05}O_3$. Inset : Comparison of field-dependent magnetization $M(H)$ of undoped and Co-doped $SrRuO_3$ at 10 K

Fig. 8(a) shows the magnetic field dependences of $\lambda_{par}$, $\lambda_{per}$, $\lambda_t$ and $\omega$ at 10 K in $SrRu_{0.95}Co_{0.05}CoO_3$. Unlike $SrRuO_3$, $\lambda_{par}$ is negative and $\lambda_{per}$ is positive at 10 K. However, both are positive at 100 K, as shown in Fig 8(b). Fig. 8(c) illustrates the temperature dependences of $\lambda_{par}$, $\lambda_{per}$, $\lambda_t$ and $\omega$ at 50 kOe obtained from isothermal field sweeps. The sign of $\lambda_{par}$ changes from negative to positive above 80 K. As a consequence, $\omega$ increases rapidly as $T_C$ is approached, before becoming negligible in the paramagnetic state. Surprisingly, the maximum value of ω near $T_C$ is comparable to that of the parent compound, even though spontaneous volume expansion in the ferromagnetic temperature is absent in the Co-doped $SrRuO_3$. We conclude that spin-orbit interaction of $Co^{2+}(d^7)$ ion contributes to the anisotropic magnetostriction below 80 K, whereas magneto-elastic coupling due to freezing of $RuO_6$ octahedral tilts/rotation in the ferromagnetic region is still preserved at higher temperatures under a magnetic field.

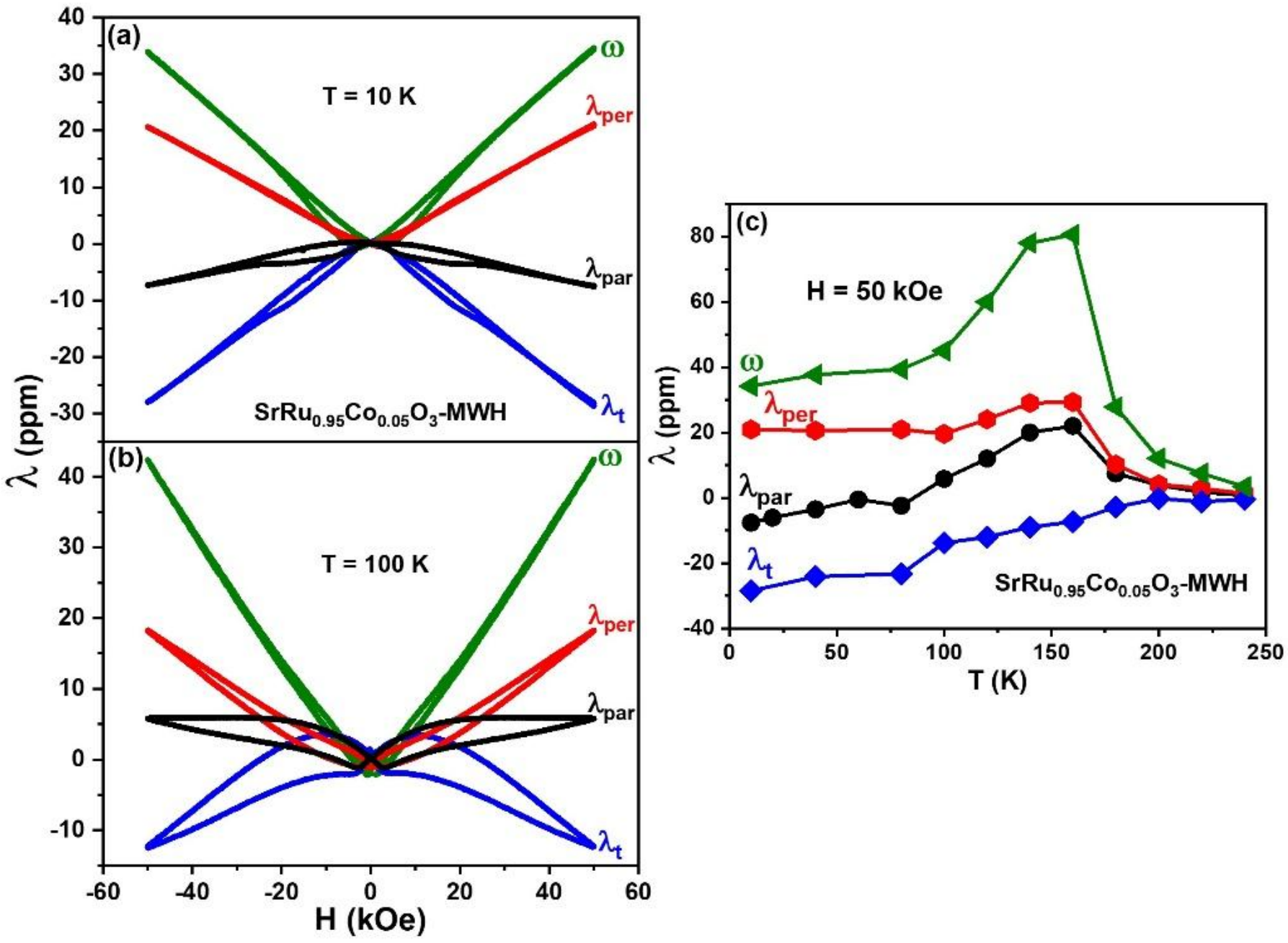

**Fig. 8.** Magnetic field dependence of parallel ($\lambda_{par}$), perpendicular ($\lambda_{per}$), volume ($\omega$) and anisotropy ($\lambda_t$) magnetostrictions of microwave-synthesized $SrRu_{0.95}Co_{0.05}O_3$ **(a)** at 10 K and **(b)** at 100 K, and **(c)** shows their temperature dependences at $H$ = 50 kOe.

It will be interesting to study doping a non-magnetic tetravalent cation such as $T^{i4+}$ ($d^0$) to weaken the magnetic coupling between neighboring $Ru^{4+}$ to investigate whether negative thermal expansion is preserved and how the field-induced positive mangetovolume effect is modified.

## 3. Summary

In summary, we successfully synthesised polycrystalline $SrRu_{1-x}Co_xO_3$ ($x$ = 0.0 and 0.05) in a very short time by microwave irradiation of oxide precursors. In zero field, thermal expansion exhibits anomalous expansion with decreasing temperature in the ferromagnetic state of $x$ = 0.0 but it is absent in $x$ = 0.05. Magnetostriction at 10 K is isotropic in $x$ = 0.0 but anisotropic in $x$ = 0.05, though it becomes isotropic above 100 K. The sample volume in the ferromagnetic state increases without saturation up to 50 kOe. The maximum volume magnetostriction in both samples is comparable (~60 ppm in $x$ = 0.0 and ~80 ppm in $x$ = 0.05), despite $x$ = 0.05 sample not showing negative thermal expansion. It is postulated that tilting of $RuO_6$ octahedra increases with magnetization in the ferromagnetic state of these Ru-based perovskites, but neutron and/or X-ray diffraction under a magnetic field is needed to understand the exact mechanism of the positive magnetovolume effect.

**ACKNOWLEDGEMENT**

R. M. acknowledges the Ministry of Education, Singapore, for supporting this work (Grants numbers: A-8000924-00-00 and A-8001933-00-00)

**DATA AVAILABILITY**

The data that support the findings of this study are available from the corresponding author upon reasonable request.